\documentclass[10pt,conference,letter]{IEEEtran}
\IEEEoverridecommandlockouts
\usepackage{epsfig}
\usepackage{amsmath}
\usepackage{amsfonts}
\usepackage{amssymb}
\usepackage{color}
\usepackage{tikz}
\usepackage{float}
\usepackage{url}
\usepackage{flushend}

\newcommand*\mycirc[1]{%
  \begin{tikzpicture}[baseline=(C.base)]
    \node[draw,circle,inner sep=1.5pt](C) {#1};
  \end{tikzpicture}}

\newcommand{\R}{\mathbb{R}}

\begin{document}
\title{Irregular Turbo Codes in Block-Fading Channels}

\author{
\authorblockN{Ghassan M. Kraidy}\thanks{G.M. Kraidy's work has been supported by the Research Council of Norway (NFR) under the project WILATI+ within the NORDITE framework.}
\authorblockA{Norwegian Univ. of Science and Technology\\
Dept. of Electronics and Telecommunications\\
7491 Trondheim, Norway\\
\tt kraidy@ieee.org}
\and
\authorblockN{Joseph J. Boutros}\thanks{J.J. Boutros's work has been supported by the
   Broadband Communications Systems project funded by Qatar Telecom (Qtel).}
\authorblockA{Texas A\&M University\\
Electrical Engineering Department\\
Education City, Doha, Qatar\\
\tt boutros@tamu.edu}
\and
\authorblockN{Albert Guill\'en i F\`abregas}
\authorblockA{University of Cambridge\\
Department of Engineering\\
Cambridge CB2 1PZ, UK\\
\tt guillen@ieee.org}
}

\maketitle
\begin{abstract}
We study  irregular binary turbo codes over non-ergodic block-fading channels. 
We first propose an extension  of channel multiplexers initially designed
for regular turbo codes. We  then show that, using these multiplexers,
irregular turbo codes that exhibit a small decoding threshold over the
ergodic Gaussian-noise channel perform very close to the outage probability
on block-fading channels, from both density evolution and finite-length perspectives.
\end{abstract}

\section{Introduction}

The  block-fading   channel  is   a  simplified  channel   model  that
characterizes  delay-constrained   communication  over  slowly-varying
fading channels \cite{ozarow1994itc,biglieri1998fci,Biglieri2005}. The
received signal at block $c$ is given by
\begin{equation}
 \boldsymbol{y}_c = \alpha_c\boldsymbol{x}_c + \boldsymbol{w}_c ~~~~~ c=1,\dotsc,n_c
\end{equation}
where       $\boldsymbol{x}_c,\boldsymbol{y},\boldsymbol{w}_c      \in
\mathbb{R}^L$  are  the  input,  output  and noise  vectors  at  block
$c=1,\dotsc,n_c$, and $L$ is the block length.  The noise components have
zero mean  and variance $N_0$,  and $\alpha_c$ is the  Rayleigh fading
coefficient  of  block $c$,  assumed  to  be  perfectly known  to  the
receiver.  Particular  instances   of  the  block-fading  channel  are
orthogonal-frequency     multiplexing     modulation    (OFDM)     and
frequency-hopping  systems,  such  as  mobile data
communications in EDGE/3G and WiMax/LTE environments.
Despite   its  simplification,  it   captures  the  essential
characteristics of delay-constrained wireless communication and yields
useful code design criteria. Since  this channel is nonergodic, it has
zero  capacity and  the fundamental  limit is  the  outage probability
\cite{ozarow1994itc,biglieri1998fci}.    It   has   been    shown   in
\cite{fabregas2006cmb}  that the  diversity  of binary  codes of  rate
$R_c$ over  an $n_c$-block  fading channel is  given by  the Singleton
bound
\begin{equation}
\delta = 1+\lfloor n_c(1-R_c)\rfloor.
\end{equation}

The design of binary linear codes for the block-fading channel has been studied in
\cite{knopp2000cbf,Malkamaki1999,fabregas2006cmb,boutros:tcd,boutros2005,boutros2009rootldpc}.
However, these binary regular codes cannot
perform closer than $1$ dB from the outage probability. 
As shown in \cite{boutros2005,boutros2009rootldpc}, the effective design procedure for
outage-approaching codes follows a two-step process:
\begin{enumerate}
\item Design block-wise maximum distance
separable (MDS) codes, that achieve the largest possible diversity given by the Singleton bound in the
block-erasure channel \cite{guillenifabregas2006cbe};
\item Reducing the decoding threshold in the AWGN channel.
\end{enumerate}

In   this   paper,   we   design   irregular binary turbo   codes
\cite{boutros2002tcd}  for   block-fading  channels.   Based   on  the
h-$\pi$-diagonal  multiplexer \cite{boutros:tcd}  we  design irregular
turbo codes  with full diversity.  We then find irregular  turbo codes
with low decoding thresholds over the AWGN. We show that the resulting
codes perform  within $0.5$ dB from the  outage probability in both density evolution and finite-length cases, achieving
the current best performance reported in the literature. 

The organization of the paper is as follows. In Section \ref{itc}, we
describe  the  structure  and  density evolution  of  irregular  turbo
codes.  The specific  block-fading  design and  density evolution  are
described  in  Section  \ref{MUX_BF}.  Section \ref{concl}  gives  the
concluding remarks.

\section{Basics on Irregular Turbo Codes}
\label{itc} 

\begin{figure}[!h]
  \begin{center} \includegraphics[width=0.8\columnwidth]{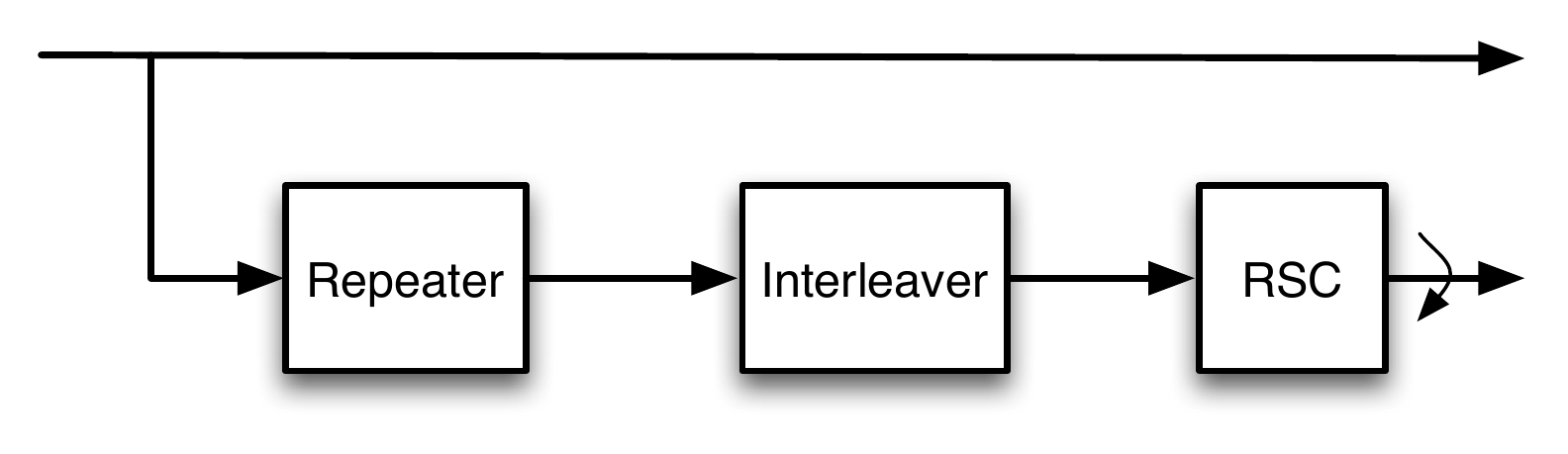}
   \caption{Systematic self-concatenated turbo encoder. Information bits are sent directly over the channel, and parity bits are generated by first repeating information bits, interleaving, and then recursive systematic convolutional (RSC) encoding. }
   \label{tc_reg}
   \end{center}
\end{figure}

In regular parallel turbo codes, the two constituent
recursive systematic convolutional (RSC) encoders are identical
({\em i.e.} same constraint length and generator polynomials)
\cite{berrou1993nsl}. This is equivalent to merging the two
constituent encoders into a single one, and doubling the size of the
interleaver.  To do so, a 2-fold repeater is added before
the interleaver $\Pi$, and we obtain a self-concatenated turbo code \cite{benedetto1998ett}\cite{loeliger1997isit} 
as shown in Fig. \ref{tc_reg}. In this representation, each information bit is connected to the code
trellis via two edges. We hence say that the
{\em degree} of the information bits is $d=2$ as shown in the propagation tree in Fig. \ref{prop_tree},
and that the turbo code is {\em regular}. 
Using this structure, one can create irregularity
by repeating a certain fraction $f_i$ of information bits $i$ times,
inducing larger protection for some bits than in the regular case \cite{frey1999allerton}.
Like for low-density parity check (LDPC) codes
\cite{richardson2001dca,luby2001ild}, irregularity can enhance the
performance of turbo codes for large block lengths
\cite{boutros2002tcd,frey1999allerton,kraidy2010itc,richardson2008mct}.  The encoder of an irregular turbo
code is similar to that of Fig. \ref{tc_reg}, with the difference
that the information bit stream is fed to a non-uniform
repeater that divides the information bits into $d$ classes with
$d=2,...,d_{\rm max}$, where $d_{\rm max}$ is the maximum bit-node
degree \cite{boutros2002tcd}.  The number of bits in a class $d$ is a fraction $f_d$ of
the total number of information bits at the turbo encoder input,
knowing that bits in class $d$ are repeated $d$ times.  Finally, the
output of the non-uniform repeater is interleaved and fed to the RSC
constituent code. In order to ensure a target rate, puncturing
is used, and only a fraction $1-f_p$ of parity bits are
transmitted, where $f_p$ is the fraction of punctured parity bits.
Now let $K$ denote the length of the information sequence, $N$ the
interleaver size, $\rho$ the rate of the RSC constituent code, and
$R_c$ the rate of the turbo code.  We have the following
\begin{equation}
\sum_{d = 2}^{d_{\rm max}} f_d = 1,~~~
\sum_{d = 2}^{d_{\rm max}} d \cdot f_d = \overline{d},
\end{equation}
\begin{equation}
N = \sum_{d = 2}^{ d_{\rm max}} d  \cdot  (f_d K) = K \cdot \overline{d},\\
\end{equation}
\begin{equation} \label{rate}
R_c = \frac{K}{K + \frac{N}{\rho} - N} = \frac{1}{1 + \left( \frac{1}{\rho} - 1\right) \overline{d}},
\end{equation}
\begin{equation}
\rho = \frac{1}{1 + \left(1 - f_p \right)\left(\frac{1}{\rho_0} - 1 \right)},
\end{equation}
\noindent
where $\rho_0 = k/n$ is the initial rate of the constituent RSC code before puncturing,
and $\overline{d}$ is the average degree of information bits. Similar to LDPC codes,
the degree distribution from an edge perspective is defined by
\begin{equation}
\lambda_d=\frac{d \cdot f_d}{\overline{d}}, ~~~~d=2 \ldots d_{\rm max}.
\end{equation}

\subsection{Density Evolution in AWGN}
We consider rate-$R_c$ irregular
turbo codes built from a rate-$\rho$ RSC constituent code and degree
profile $\lbrace f_d \rbrace_{d=2,...,d_{\rm max}}$.  Due to the symmetry of the channel, we assume that the
all-zero codeword is modulated into $\boldsymbol{x}={+1,+1,...,+1}$ and
transmitted over an AWGN channel with noise variance~$N_0$.  
At the channel output, each received sample can be written as
$y=x+w=1+w$, so the log-likelihood ratio (LLR) is
given by the well-known expression:
\begin{equation}
\label{equ_chan_message}
\mathcal{M}_0 = \log \frac{p(y|x=+1)}{p(y|x=-1)} = \frac{2}{N_0}y =
\frac{2}{N_0}(1+w).
\end{equation}
\noindent 
We have $\mathcal{M}_0 \sim \mathcal{N}(\frac{2}{N_0}, \frac{4}{N_0})$,
the associated probability density function will be denoted by $p_0(x)$.

\begin{figure}[!h]
\begin{center}
   \includegraphics[width=0.9\columnwidth]{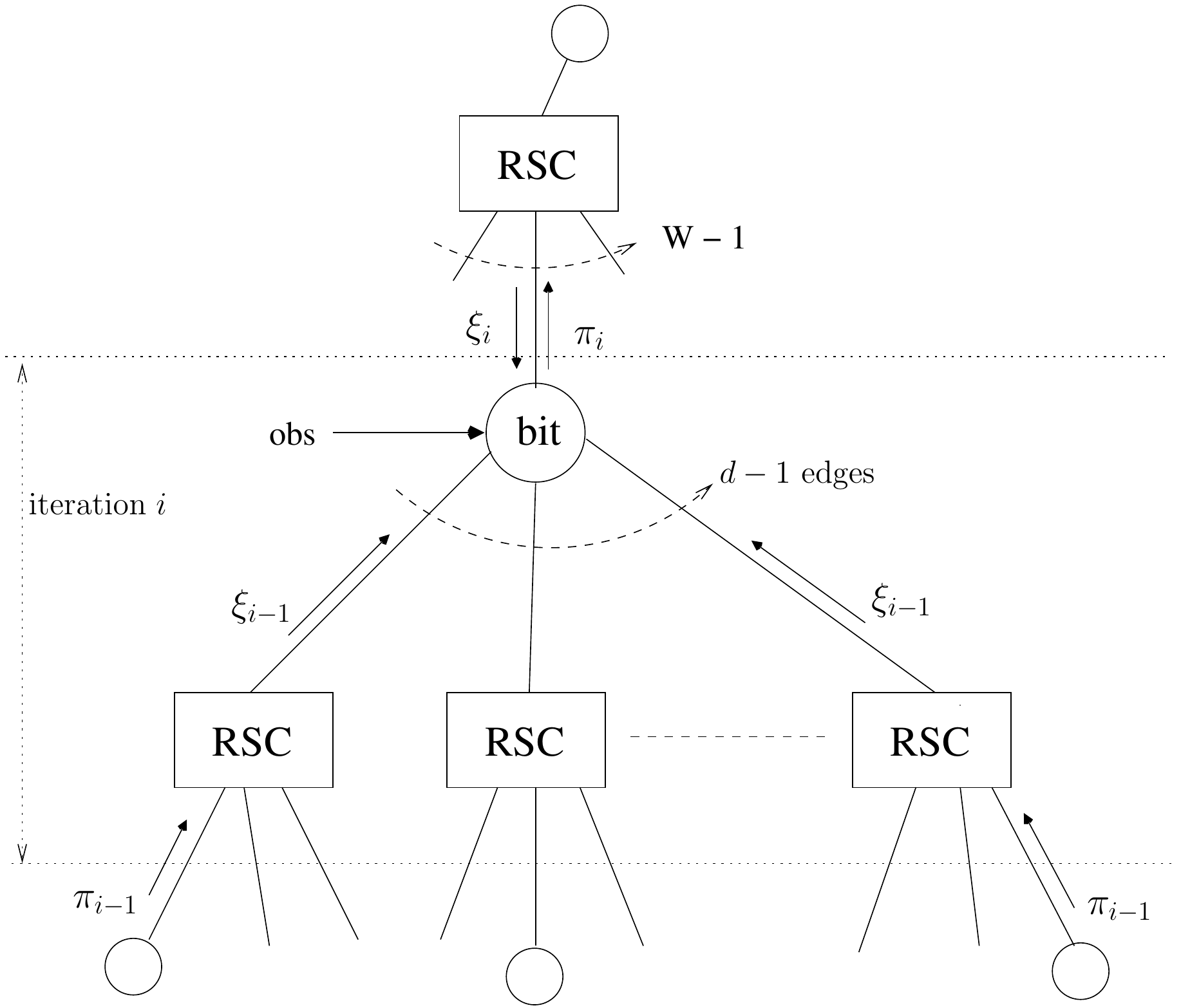}
   \caption{Propagation tree used in density evolution for an irregular turbo code. The $\pi_i$ represents {\em a priori} probability, and the $\xi_i$ the extrinsic probability. Circles represent bitnodes, and rectangles are local neighborhood RSC trellis constraints.}
   \label{prop_tree}
\end{center}
\end{figure}

The local neighborhood tree  for  an information bit belonging to an 
acyclic asymptotically large irregular  turbo code  is  shown in  Fig.
\ref{prop_tree}.   The index $i$ refers to the decoding iteration number.
A   bitnode  of   degree   $d$ has  $d-1$ incoming extrinsic probabilities
$\xi_i$ and one outgoing {\em a priori} probability $\pi_{d,i}$ 
which also plays the role of a partial {\em a
posteriori}  probability (APP). The total APP may be obtained
by combining $\pi_{d,i}$ with an extra extrinsic probability.
The message associated to $\xi_i$ is
$\mathcal{M}_i = \frac{\log(\xi_i({\rm bit}=0))}{\log(\xi_i({\rm bit}=1))}$
and its probability density function is $p_{\mathcal{M}_i}(x)$.
Given $d$ and $i$, the probability density function
of log-ratio messages associated to $\pi_{d,i}$ will be denoted by $p_{d,i}(x)$.
Following \cite{richardson2008mct} we have that
\begin{equation}
\label{obs} p_{d,i}(x) = \mathcal{F}^{-1}\left[\mathcal{F}\left[
p_0(x)\right] \mathcal{F}^{d-1} \left[ p_{\mathcal{M}_i}(x)\right]\right]
\end{equation}
where $\mathcal{F}$ denotes the Fourier  transform operator.
Based on partial a posteriori probabilities, the average bit error probability at iteration $i$ is defined as
\begin{equation}
 P_b(i) = \sum_{d=2}^{d_{\rm max}} f_d P_b(d,i)
\end{equation}
where $P_b(d,i)$ is the bit error probability of class $d$ given by
the area under the tail of $p_{d,i}(x)$.

At an RSC checknode level as illustrated in  Fig. \ref{prop_tree}, 
based on a priori input $\pi_{i-1}$ with pdf $p_{i-1}(x)$,
an accurate estimation of $p_{\mathcal{M}_i}(x)$ is made via
a forward-backward algorithm \cite{bcjr}  applied on a sufficiently large trellis window 
of size $W$ centered around the information bit. Since we are dealing with random ensembles
of irregular turbo codes, we have
\begin{equation}
p_i(x)= \sum_{d=2}^{d_{\rm max}} \lambda_d ~p_{d,i}(x).
\end{equation}

Given an irregular turbo ensemble, its decoding {\em threshold} is the minimal signal-to-noise ratio $E_b/N_0$ 
for which $P_b(i)$ vanishes with $i$. The threshold can be determined via Density Evolution (DE) \cite{richardson2008mct},
a procedure where $p_i(x)$ is updated from $p_{i-1}(x)$ by propagating probabilistic densities
through the tree graph of Fig. \ref{prop_tree}.

\subsection{Numerical results for AWGN}

The DE method gives the limiting convergence behavior of capacity-approaching  codes, 
and it is used to
find     optimal    degree    profiles     for    LDPC     codes    in
\cite{richardson2001dca,richardson2008mct,chung2001dld}.   By  setting
the  average  degree  to be  $\overline{d}  =  3$  and using  the  RSC
$(13,15)_8$ constituent code, we obtained powerful half-rate irregular
turbo  codes with  different degree  profiles; for  example  by taking
$f_2=0.9$, $f_9=0.04$,  and $f_{15}=0.06$, the threshold  is $0.31$ dB.  The  distributions $f_2=0.923$ and  $f_{15}=0.077$ or $f_2=0.9$
and $f_{12}=0.1$, yield  a $0.36$ dB threshold. Recall  that Shannon
limit  for half-rate  coding over  the AWGN  channel  is approximately
$0.18$  dB.  The  irregular   turbo  code  defined  by  $f_2=0.9$  and
$f_{12}=0.1$  is   used  later   in  section  \ref{MUX_BF}   over  the
block-fading channel.

\section{Irregular Turbo Codes over Block-Fading Channels \label{MUX_BF}} 
In \cite{boutros:tcd}, the authors proposed
multiplexer design for regular parallel turbo codes that ensure full
diversity and optimal coding gain. However, as the self-concatenated
structure of the code involves only one constituent code, the
generalization of the so-called h-$\pi$-diagonal multiplexers initially designed for regular parallel
turbo codes is not straightforward. Without loss of generality, we restrict our design to irregular
turbo codes over block-fading channels with $n_c=2$ blocks and rate $R_c=1/2$. The
extension to block-fading channels with more fading blocks follows similar arguments
but it is not discussed in this paper. Special care should be taken
when designing a turbo code that achieves the Singleton bound 
without attaining full diversity, i.e., $n_c > \delta \ge 1/R_c$.

In an irregular turbo code with
average degree $\overline{d}$, a bit is connected to the trellis of
the code via $\overline{d}$ edges on average. Following the identity
$N=K \overline{d}$, this can be seen as a
``parallel'' turbo code with $\beta$ constituent codes, where:
\begin{equation}
 \beta=\lceil ~ \overline{d}~ \rceil
\end{equation}
In order to achieve high coding gains, the h-$\pi$-diagonal
multiplexer should be extended to irregular turbo
codes. We consider constituent RSC
codes with initial coding rate $\rho_0~=~1/2$. To keep the
structure of the multiplexer, only half of the parity bits of the
first RSC constituent code should be punctured, knowing that the overall
rate $R_c$ should remain fixed. Now let $\phi_p$ be the fraction of
parity bits to be punctured from every RSC constituent code starting
from the second one. We have that:
\begin{equation}
\phi_p = \frac{\beta f_p - \frac{1}{2}}{\beta - 1}
\end{equation}
The general h-$\pi$-diagonal multiplexer is shown in Fig.
\ref{irreg_mux_1}, where $b$ is the information bit, and $s_j$ is
the parity bit of constituent code $j$.
As an example, we consider a half-rate irregular turbo code with
$\beta = \overline{d} = 3$. This gives $f_p=0.66$ and $\phi_p=0.75$,
so 3 parity bits out of 4 are punctured from both RSC 2 and RSC 3. 
Again, we consider a half-rate irregular turbo code with
$\overline{d} = 2.727$. We get $f_p=0.63$ and $\phi_p=0.7$. The
puncturing pattern is then slightly different from that of the
previous example, as in a period of length 20, there is one more bit that is 
sent over the channel.\\

\begin{figure}[!h]
\begin{center}
\small{
\begin{center}
\begin{tabular}{|c|c|c|c|c|c|}
\hline RSC 1 (information) & $b$             & \mbox{  } 1 & 2 & 1 & 2 \\
\hline RSC 1 (parity) & $s_1$           & \mbox{  } 2 & X & 2 & X \\
\hline RSC 2 (parity) & $\pi^{-1}(s_2)$ & \mbox{ }  X & 1/X &  X & 1/X\\
\hline \vdots & \vdots & \vdots & \vdots & \vdots & \vdots\\
\hline RSC $\beta$ (parity) & $\pi^{-1}(s_{\beta})$ & \mbox{ } X & 1/X &  X & 1/X\\
\hline
\end{tabular}
\end{center}
}
\end{center}
\caption{H-$\pi$-diagonal multiplexer of a half-rate irregular turbo code
built from $\rho_0 = 1/2$ constituent RSC code. The number of rows is $\beta+1$ where $\beta=\lceil ~\bar{d}~ \rceil$. 
One parity bit out of two is punctured from RSC 1. There is a
fraction $\phi_p$ of punctured parity bits per row (represented by
an X) starting from RSC 2.} \label{irreg_mux_1}
\end{figure}

\subsection{Density evolution on BF channel}
\label{DE_SEC}

In this section we study the word error rate performance of half-rate
irregular turbo codes over a two-state block-fading channel via
density evolution. As with the AWGN channel, we assume that the all-zero codeword is modulated
into $x={+1,+1,...,+1}$ and transmitted over a block-fading channel
with $n_c$ states ($n_c=2$ in our case).

For a given fading instance $\boldsymbol{\alpha}=(\alpha_1,\alpha_2)$, the irregular turbo code
ensemble is observing two types of channel messages, $\mathcal{M}_{0,1} \sim \mathcal{N}(\frac{2\alpha_1}{N_0}, \frac{4\alpha_1^2}{N_0})$
and $\mathcal{M}_{0,2} \sim \mathcal{N}(\frac{2\alpha_2}{N_0}, \frac{4\alpha_2^2}{N_0})$, as in (\ref{equ_chan_message}).
DE is performed in a similar fashion as described in Section \ref{itc}.A after taking into account
the multiplexing of bits (i.e. which channel assigned to which bit) as defined in Fig. \ref{irreg_mux_1}.
At a fixed SNR, it is possible to determine via DE whether the average bit error probability $P_b(i)$ vanishes with $i$ or not.
When $P_b(i) \nrightarrow 0$ as $i \rightarrow +\infty$, we say that a density evolution outage (DEO) occurs.

Now, let us define the following indicator function:
\begin{equation}
\mathbf{1}_{\rm DEO}(\alpha) = 
\left\lbrace 
\begin{array}{l}
0, ~~P_b(i) \rightarrow 0,\\
~\\
1, ~~P_b(i) \nrightarrow 0.
\end{array}
\right.
\end{equation}
The probability of a DEO is then given by
\begin{equation}
P_{\rm DEO} = \int_{\boldsymbol{\alpha} \in \R^2} \mathbf{1}_{\rm DEO}(\boldsymbol{\alpha}) p(\boldsymbol{\alpha}) d\boldsymbol{\alpha} 
= \int_{\boldsymbol{\alpha} \in V_o} p(\boldsymbol{\alpha})d\boldsymbol{\alpha},
\end{equation}
where $V_o$ is the outage region for the irregular turbo code ensemble under DE, i.e., 
\begin{equation}
V_o = \left\lbrace \boldsymbol{\alpha} \in \mathbb{R}_+^{n_c} \mid  \mathbf{1}_{\rm DEO}(\boldsymbol{\alpha})=1 \right\rbrace.
\end{equation}
The $(n_c-1)$-dimensional surface separating $V_o$ from its complementary in $\mathbb{R}_+^{n_c}$ is called the outage boundary.
Thus, DE on a block-fading channel is a method to determine the outage boundary for a given
turbo code ensemble at a given SNR. The information-theoretical boundary related to the outage probability
is defined by the equality $C(\boldsymbol{\alpha}, E_b/N_0)=R_c$, where $C$ is the channel capacity (or mutual information)
under a certain type of input alphabet.

For an infinite-length code ensemble, it is easy to show that the word error probability $P_{ew}$ satisfies \cite{boutros2009rootldpc}
\begin{equation}
\label{equ_Pew_lower_bound}
P_{\rm DEO} \le P_{ew}.
\end{equation}
Consequently, the outage probability found by DE is a lower bound for the word error probability
and can be compared to the information outage probability. Equality in (\ref{equ_Pew_lower_bound}) occurs if the block threshold is equal
to the bit threshold \cite{jin2005bei}.

\subsection{Numerical results on BF channel}
\begin{figure}[!h]
\begin{center}
\includegraphics[height=0.99\columnwidth]{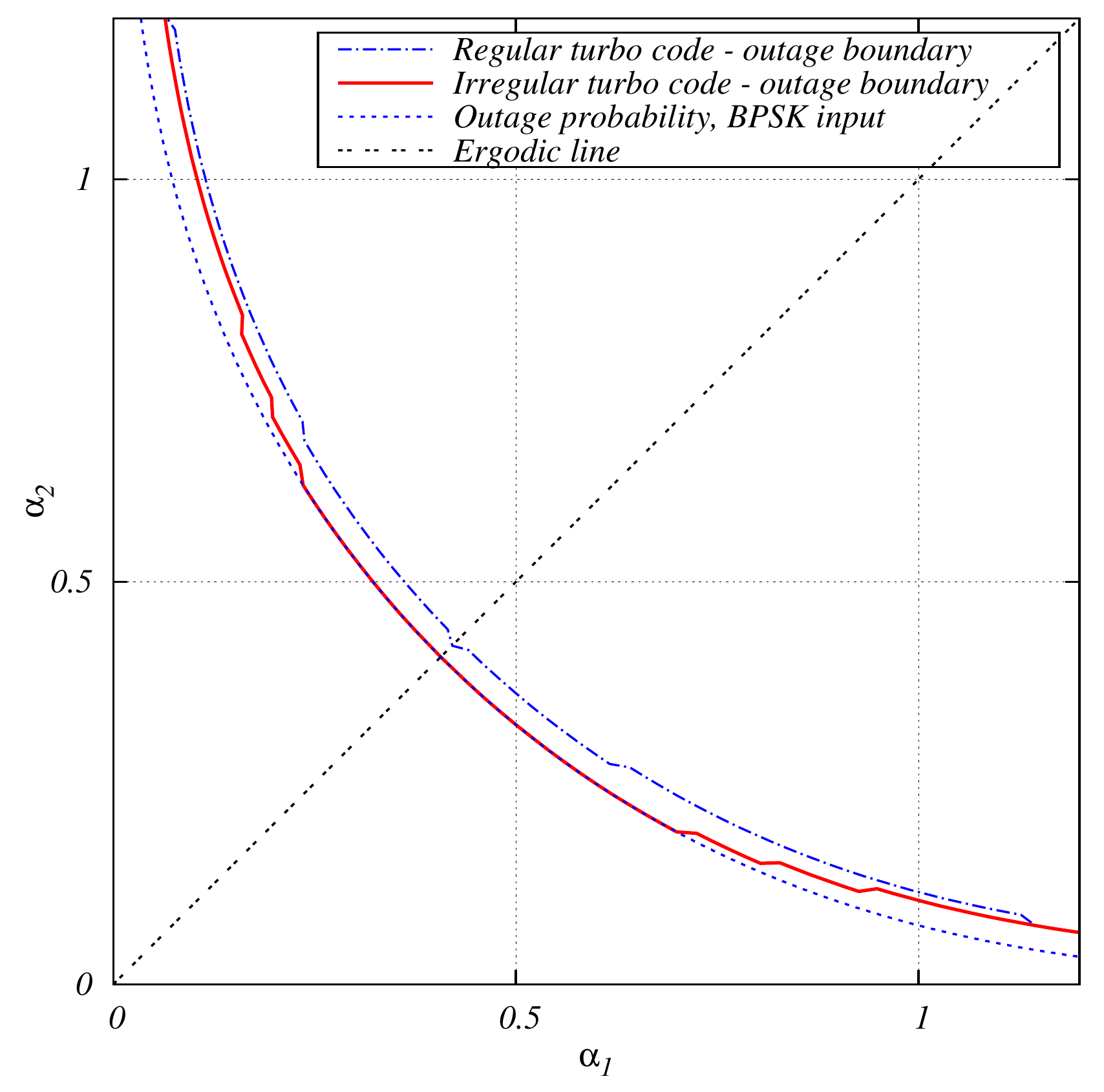}
\caption{Outage boundary of regular and irregular turbo codes under h-$\pi$-diagonal multiplexing and with the RSC $(13,15)_8$ constituent code at $E_b/N_0=8$dB. Circles filled with crosses correspond to the fading pairs in which irregular turbo codes outperform regular codes. Although the two codes have similar performance with largely unbalanced fading pairs, the irregular code outperforms the regular code in the vicinity of the ergodic line. } \label{boundary}
\end{center}
\end{figure}

Fig.  \ref{boundary}  compares  the  outage boundary  of  regular  and
irregular  turbo codes with  the $8$-state  RSC$(13,15)_8$ constituent
code and h-$\pi$-diagonal multiplexing at $E_b/N_0=8$dB. The irregular
turbo code is the best one from Section \ref{DE_SEC}, with a threshold
of  $0.31$dB on  the  AWGN  channel. The  boundaries  are computed  by
picking points orthogonal to the BPSK input outage. Although irregular
and  regular codes  have  similar performance  for largely  unbalanced
fading  pairs,  the  irregular  turbo  code  performs  better  in  the
neighborhood  of the ergodic  line.  It  actually approaches  the BPSK
input    outage    border   over    a    large    range   of    fading
pairs. 

\begin{figure}[!h]
\vspace{-0.6cm}
\hspace{-1cm}\includegraphics[width=1.15\columnwidth]{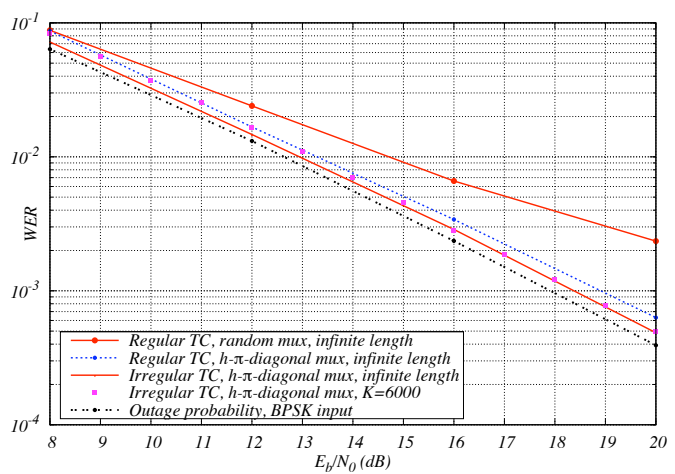}
\vspace{-1.5cm}
\begin{center}
\caption{Word error rate for
$R_c=\frac{1}{2}$ turbo codes over the block-fading channel with
$n_c=2$,  RSC $(13,15)_8$ constituent code and BPSK modulation. Performance of codes is invariant with codeword length, and it was estimated using both the density evolution algorithm and Monte Carlo simulations.} \label{comparison_DE}
\end{center}
\end{figure}

Fig. \ref{comparison_DE} shows  the word error rate performance
of the same codes and h-$\pi$-diagonal multiplexing under both density
evolution and Monte  Carlo simulation with $K=6000$  bits. As we observe, both DE performance and finite-length are very close to the outage probability (within 0.5 dB). Note that, as observed in \cite{fabregas2006cmb,boutros:tcd,boutros2005,boutros2009rootldpc}, irregular turbo codes are good for the block-fading channel, in the sense that their performance is insensitive to the block length.

For finite length simulations,  the repeater should
be designed in a special way, as shown in Fig. \ref{irreg_mux_3}. Bits
are  divided into two  groups, and  only the  information bits  of the
first  RSC  are  transmitted   over  the  channel:  circled  bits  are
transmitted over the $1^{st}$  channel state, and non-circled bits are
sent over  the $2^{nd}$ channel  state.  To guarantee  full diversity,
the  decoder should always  find its  way through  the trellis  of the
code, thus  bits corresponding to  the same trellis  transition should
not be sent  over the same channel state  \cite{boutros:tcd}. In order
to ensure this  property, bits of degree greater than  2 are placed in
the       H      positions       in      the       multiplexer      of
Fig. \ref{irreg_mux_3}. Repetition  is thus done in a  way that if the
2nd channel  state is unreliable,  decoding can be  successful through
RSC 2 and RSC 3, and if the 1st channel state is unreliable, RSC 1 can
decode  the  received codeword.  

\begin{figure*}[htpb!]
\begin{center}
\footnotesize{
\begin{center}
\begin{tabular}{|c|cccccc||cccccc||cccccc|}
\hline  & \mbox{  } &  &  & RSC 1 & &  &   &     &  & RSC 2 &  &  &  &  &   & RSC 3 &  &  \\
\hline $\text{I}$          & \mbox{  } \mycirc{1} & {2} & \mycirc{3} & {4} & \mycirc{5} & {6} & \mycirc{1}  & {2} & \mycirc{H} & \mycirc{H} & \mycirc{5}  & {6} & \mycirc{H} & \mycirc{H} & \mycirc{3}  & {4} & \mycirc{H} & \mycirc{H} \\
\hline $\text{P}$          & \mbox{  } {$p_1$} & X &{$p_3$} & X & {$p_5$} & X & X & \mycirc{$p_2$} & X & X & X & \mycirc{$p_6$} & X & X & X & \mycirc{$p_4$} & X & X \\
\hline
\end{tabular}
\end{center}
}
\end{center}
\caption{H-$\pi$-diagonal multiplexer of a half-rate irregular turbo
code with $\overline{d}=3$ transmitted on a $2$-state
block-fading channel using a punctured half-rate constituent RSC
code. The irregular turbo encoder is built using 3 constituent encoders, where only the information bits (on the line labeled with I) of RSC 1 are transmitted over the channel. The bits $p_i$ correspond to parity bits, the X represents punctured parity bits, and the bits labeled H correspond to bits with degree higher than 2. The circled bits are sent over the the $1^{st}$ channel state, the other bits are sent over the $2^{nd}$ state. In order to achieve full diversity, some of the circled information bits should be repeated more than twice and fed to RSC 2 and RSC 3. } \label{irreg_mux_3}
\end{figure*}

Note that although the density
evolution convergence criterion is  based on bit error probability, it
is relevant  to assume  that the word  error probability  of irregular
turbo codes has an equivalent decoding threshold under density evolution.  
In fact, it
was shown in \cite{jin2005bei} that the word and bit error probability
of   certain  LDPC  codes,   among  which   the  class   of  Irregular
Repeat-Accumulate  (IRA)   codes  \cite{jin2000ira},  have identical thresholds.
Irregular  turbo codes  can  be seen  as  IRA codes  that are  decoded
iteratively  using  a  different  scheduling, that  results  from  the
difference between forward-backward and belief-propagation decoding.

\section{Conclusions}
\label{concl} In this paper, we presented irregular turbo codes that
are capable of closely approaching the outage probability of the
block-fading channel, both in terms of density evolution (infinite length) and finite length. The design method is based on two steps. First, a
suitable full-diversity multiplexer was designed. Second, codes were
optimized over the AWGN channel through density evolution. This
represents the best family of codes over
the block-fading channel reported in the literature. \vspace{-2mm}

\bibliographystyle{IEEE}
\bibliography{refs}

\end{document}